\documentclass[conference,letterpaper]{IEEEtran}
\IEEEoverridecommandlockouts%
\usepackage{cite}
\usepackage[para]{footmisc}
\usepackage{amsmath,amssymb,amsfonts}
\usepackage{algorithmic}
\usepackage{graphicx}
\graphicspath{{./img/}}
\usepackage{textcomp}
\usepackage[table,xcdraw]{xcolor}
\usepackage{diagbox}
\usepackage{colortbl}
\usepackage{todonotes}
\usepackage{multirow}
\usepackage{url}
\usepackage{nohyperref} %%For export activate for no hyperlinks.
\usepackage[per-mode=symbol,group-separator={,},group-minimum-digits={3}]{siunitx}
\usepackage{caption}
\usepackage{subcaption}
\usepackage[nohyperlinks,nolist]{acronym} %\ac{} ausgabe, \acl{} langausgabe, \acf{} zweites erstes mal, \acp plural
\usepackage[]{algorithm2e}
\usepackage{tikz}
\usepackage{tikz-layers}
\usetikzlibrary{calc}
\usepackage{tabularx}
\newcolumntype{Y}{>{\centering\arraybackslash}X}
\newcolumntype{L}{>{\arraybackslash}X}
\newcolumntype{R}{>{\raggedleft\arraybackslash}X}
\newcolumntype{C}[1]{>{\centering\arraybackslash}p{#1}}
\usepackage{booktabs}
\usepackage[absolute,showboxes]{textpos}
\usetikzlibrary{patterns}
\usetikzlibrary{spy}
\usetikzlibrary{arrows.meta}
\usetikzlibrary{fit}
\usepackage{pgfplots}
\pgfplotsset{compat=newest}
\usepgfplotslibrary{units}
\usepgfplotslibrary{groupplots}
\usepgfplotslibrary{statistics}
\makeatletter
\pgfplotsset{
 unit code/.code 2 args=
   \begingroup
   \protected@edef\x{\endgroup\si{#2}}\x
}
\makeatother
\usetikzlibrary{matrix}
\usetikzlibrary{shapes.misc}

\definecolor{CoreGray}{HTML}{BFBFBF}
\definecolor{CoreBlack}{HTML}{333333}
\definecolor{CoreBlue}{HTML}{002E7D}
\definecolor{CoreGreen}{HTML}{6AAC8E}
\definecolor{CoreRed}{HTML}{C80000}
\definecolor{CoreYellow}{HTML}{E6AC00}
\definecolor{CoreWhite}{HTML}{FFFFFF}
\definecolor{CoreMiddleGray}{HTML}{e5e5e5}
\colorlet{LightCoreGray}{CoreGray!20}
\colorlet{LightCoreBlack}{CoreBlack!20}
\definecolor{CoreLightBlue}{HTML}{8BBDEB}
\colorlet{LightCoreGreen}{CoreGreen!30}
\colorlet{LightCoreRed}{CoreRed!20}
\colorlet{LightCoreYellow}{CoreYellow!20}
\colorlet{LightCoreWhite}{CoreWhite!20}

\newcommand\codeword[1]{\mbox{\texttt{\textcolor{CoreBlack}{#1}}}}

\usepackage[absolute,showboxes]{textpos}

\makeatletter
\newcommand\footnoteref[1]{\protected@xdef\@thefnmark{\ref{#1}}\@footnotemark}
\makeatother

\begin{document}

% Automatically shrink down the bib entries:
\bstctlcite{my:BSTcontrol}

\title{
    Authenticated and Secure Automotive Service Discovery with DNSSEC and DANE
    % \thanks{}
}
\author{\IEEEauthorblockN{Mehmet Mueller}\IEEEauthorblockA{\href{http://www.haw-hamburg.de/ti-i}{\textit{Dept. Computer Science}},
\href{http://www.haw-hamburg.de/ti-i}{\textit{Hamburg University of Applied Sciences}}, Germany \\
\{\href{mailto:mehmet.mueller@haw-hamburg.de}{mehmet.mueller}\}@haw-hamburg.de}
}
\author{\IEEEauthorblockN{Mehmet Mueller, Timo H\"ackel, Philipp Meyer, Franz Korf, and Thomas C. Schmidt}%
\IEEEauthorblockA{\href{http://www.haw-hamburg.de/ti-i}{\textit{Dept. Computer Science}},
\href{http://www.haw-hamburg.de/ti-i}{\textit{Hamburg University of Applied Sciences}}, Germany \\
\{\href{mailto:mehmet.mueller@haw-hamburg.de}{mehmet.mueller}, %
\href{mailto:timo.haeckel@haw-hamburg.de}{timo.haeckel}, %
\href{mailto:philipp.meyer@haw-hamburg.de}{philipp.meyer}, %
\href{mailto:franz.korf@haw-hamburg.de}{franz.korf}, %
\href{mailto:t.schmidt@haw-hamburg.de}{t.schmidt}\}@haw-hamburg.de} %
}

\maketitle

\setlength{\TPHorizModule}{\paperwidth}
\setlength{\TPVertModule}{\paperheight}
\TPMargin{5pt}
\begin{textblock}{0.8}(0.1,0.02)
     \noindent
     \footnotesize
     If you cite this paper, please use the original reference:
     Mehmet Mueller, Timo H\"ackel, Philipp Meyer, Franz Korf, and Thomas C. Schmidt. 
     ``Authenticated and Secure Automotive Service Discovery with DNSSEC and DANE,'' In: \emph{Proceedings of the 14th IEEE Vehicular Networking Conference (VNC)}. IEEE, April 2023.
\end{textblock}

\begin{abstract}
    Automotive softwarization is progressing and future cars are expected to operate a \acl{SOA} on multipurpose compute units, which are interconnected via a high-speed Ethernet backbone. The AUTOSAR architecture foresees a universal middleware called \acs{SOME/IP} that provides the service primitives, interfaces, and application  protocols on top of Ethernet and IP. \acs{SOME/IP} lacks a robust security architecture, even though security is an essential in future Internet-connected vehicles. 
	In this paper, we augment the \acs{SOME/IP} service discovery with an authentication and certificate management scheme based on DNSSEC and DANE. We argue that the deployment of well-proven, widely tested standard protocols should serve as an appropriate basis for a robust and reliable security infrastructure in cars. Our solution enables on-demand service authentication in offline scenarios, easy online updates, and remains free of attestation collisions. We evaluate our extension of the common \textit{vsomeip} stack and find performance values that fully comply with car operations. 
\end{abstract}


%%%% 	Acronyms    %%%%
% !TEX root = ../main.tex
\begin{acronym}
	% A
	\acro{ACC}[ACC]{Adaptive Cruise Control}
	\acro{ACDC}[ACDC]{Automotive Cyber Defense Center}
	\acro{ACL}[ACL]{Access Control List}
	\acro{ADS}[ADS]{Anomaly Detection System}
	\acroplural{ADS}[ADSs]{Anomaly Detection Systems}
	\acro{ADAS}[ADAS]{Advanced Driver Assistance Systems}
	\acro{API}[API]{Application Programming Interface}
	\acro{AVB}[AVB]{Audio Video Bridging}
	\acro{ARP}[ARP]{Address Resolution Protocol}
	% B
	\acro{BE}[BE]{Best-Effort}
	% C
	\acro{CA}[CA]{Certification Authority}
	\acro{CAN}[CAN]{Controller Area Network}
	\acro{CBM}[CBM]{Credit Based Metering}
	\acro{CBS}[CBS]{Credit Based Shaping}
	\acro{CNC}[CNC]{Central Network Controller}
	\acro{CMI}[CMI]{Class Measurement Interval}
	\acro{CoRE}[CoRE]{Communication over Realtime Ethernet}
	\acro{CRA}[CRA]{Cyber Resilience Act}
	\acro{CT}[CT]{Cross Traffic}
	\acro{CM}[CM]{Communication Matrix}
	% D
	\acro{DANE}[DANE]{DNS-Based Authentication of Named Entities}
	\acro{DDoS}[DDoS]{Distributed Denial of Service}
	\acro{DDS}[DDS]{Data Distribution Service}
	\acro{DCPS}[DCPS]{Data-Centric Publish-Subscribe}
	\acro{DNS}[DNS]{Domain Name System}
	\acro{DNSSEC}[DNSSEC]{Domain Name System Security Extensions}
	\acro{DoS}[DoS]{Denial of Service}
	\acro{DPI}[DPI]{Deep Packet Inspection}
	% E
	\acro{E/E}[E/E]{Electrical/Electronic}
	\acro{ECU}[ECU]{Electronic Control Unit}
	\acroplural{ECU}[ECUs]{Electronic Control Units}
	% F
	\acro{FDTI}[FDTI]{Fault Detection Time Interval}
	\acro{FHTI}[FHTI]{Fault Handling Time Interval}
	\acro{FRTI}[FRTI]{Fault Reaction Time Interval}
	\acro{FTTI}[FTTI]{Fault Tolerant Time Interval}
	% G
	\acro{GCL}[GCL]{Gate Control List}
	% H
	\acro{HTTP}[HTTP]{Hypertext Transfer Protocol}
	\acro{HMI}[HMI]{Human-Machine Interface}
	\acro{HPC}[HPC]{High-Performance Controller}
	% I
	\acro{IA}[IA]{Industrial Automation}
	\acro{IDS}[IDS]{Intrusion Detection System}
	\acroplural{IDS}[IDSs]{Intrusion Detection Systems}
	\acro{IEEE}[IEEE]{Institute of Electrical and Electronics Engineers}
	\acro{IETF}[IETF]{Internet Engineering Task Force}
	\acro{IoT}[IoT]{Internet of Things}
	\acro{IP}[IP]{Internet Protocol}
	\acro{ICT}[ICT]{Information and Communication Technology}
	\acro{IVNg}[IVN]{In-Vehicle Networking}
	\acro{IVN}[IVN]{In-Vehicle Network}
	\acroplural{IVN}[IVNs]{In-Vehicle Networks}
	%J
	% L
	\acro{LIN}[LIN]{Local Interconnect Network}
	% M
	\acro{MAC}[MAC]{Message Authentication Code}
	\acro{MITM}[MITM]{Man-In-The-Middle}
	\acro{MOST}[MOST]{Media Oriented System Transport}
	% N
	\acro{NADS}[NADS]{Network Anomaly Detection System}
	\acroplural{NADS}[NADSs]{Network Anomaly Detection Systems}
	% O
	\acro{OEM}[OEM]{Original Equipment Manufacturer}
	\acro{OMG}[OMG]{Object Management Group}
	\acro{OTA}[OTA]{Over-the-Air}
	%P
	\acro{P4}[P4]{Programming Protocol-independent Packet Processors}
	\acro{PCP}[PCP]{Priority Code Point}
	\acro{PKI}[PKI]{Public Key Infrastructure}
	% R
	\acro{RC}[RC]{Rate-Constrained}
	\acro{REST}[ReST]{Representational State Transfer}
	\acro{RPC}[RPC]{Remote Procedure Call}
	% S
	\acro{SD}[SD]{Service Discovery}
	\acro{SDN}[SDN]{Software-Defined Networking}
	\acro{SDN4CoRE}[SDN4CoRE]{Software-Defined Networking for Communication over Real-Time Ethernet}
	\acro{SIEM}[SIEM]{Security Information and Event Management}
	\acro{SOA}[SOA]{Service-Oriented Architecture}
	\acro{SOC}[SOC]{Security Operation Center}
	\acro{SOME/IP}[SOME/IP]{\textit{Scalable service-Oriented MiddlewarE over IP}}
	\acro{SPOF}[SPOF]{Single Point of Failure}
	\acro{SR}[SR]{Stream Reservation}
	\acro{SRP}[SRP]{Stream Reservation Protocol}
	\acro{SW}[SW]{Switch}
	\acroplural{SW}[SW]{Switches}
	% T
	\acro{TAS}[TAS]{Time-Aware Shaping}
	\acro{TCP}[TCP]{Transmission Control Protocol}
	\acro{TDMA}[TDMA]{Time Division Multiple Access}
	\acro{TLS}[TLS]{Transport Layer Security}
	\acro{TSN}[TSN]{Time-Sensitive Networking}
	\acro{TSSDN}[TSSDN]{Time-Sensitive Software-Defined Networking}
	\acro{TT}[TT]{Time-Triggered}
	\acro{TTE}[TTE]{Time-Triggered Ethernet}
	% U
	\acro{UDP}[UDP]{User Datagram Protocol}
	\acro{UN}[UN]{United Nations}
	% Q
	\acro{QoS}[QoS]{Quality-of-Service}
	% V
	\acro{V2X}[V2X]{Vehicle-to-X}
	%W
	\acro{WS}[WS]{Web Services}
	% Z
	\acro{ZC}[ZC]{Zone Controller}

\end{acronym}

%%%%	Document    %%%%
%!TEX root = ../main.tex

\section{Introduction}%
\label{sec:introduction}

Future cars will  connect to the Internet as well as to other vehicles and infrastructure  (\ac{V2X}) for improving road safety, traffic efficiency, and driver comfort.
This opens a large attack surface across communication interfaces~\cite{cmkas-ceaas-11,mv-sraas-14,mv-reupv-15} and in-car software~\cite{kpk-essoe-19,xllzl-ssdai-22}.
Nevertheless, current automotive protocols and \acp{ECU} often lack security mechanisms~\cite{mrfmsvcJR21} since they were designed for a closed environment.
Industry standards (e.g., ISO/SAE 21434~\cite{iso-sae-21434}) and legislation (e.g., the European \acl{CRA}) demand automotive security throughout the entire supply chain for hardware and software.

\ac{SOA} for automotive software emerges as a paradigm that facilitates service provisioning by various suppliers of an \ac{OEM}. 
\ac{SOME/IP}~\cite{aspsJR21} -- standardized by AUTOSAR -- is the most widely deployed middleware tailored to the automotive environment and implements service-oriented communication via IP and Automotive Ethernet~\cite{mk-ae-15}. 
Paired with \ac{TSN}~\cite{ieee8021q-18}, Automotive Ethernet can meet real-time requirements. 
In this architecture, services are envisioned to be dynamically updated and orchestrated on the vehicle \acp{ECU}~\cite{kakww-dsosa-19}.
Therefor \ac{SOME/IP} provides a complementary \ac{SD}~\cite{assdJR21} that detects service availability and establishes sessions between producers and consumers.
\ac{SOME/IP}, however, does not verify the authenticity of service providers.

The problem of securing \ac{SD} is not unique to the automotive domain.
On the Internet, the endpoints of services are determined with the help of the \ac{DNS}. 
Its  \ac{DNSSEC}~\cite{RFC-2535} ensure data integrity and authenticity of the \ac{DNS} records. 
In addition, \ac{DANE}~\cite{RFC-6698} binds public certificates to names to ensure the authenticity of the connection endpoint unambiguously and without attestation collisions.
\ac{DNSSEC} and \ac{DANE} are well-established and widely deployed   Internet standards with almost eight million \ac{DNSSEC} verified zones and more than half a million \ac{DANE} enabled zones on the Internet\footnote{SecSpider Global DNSSEC deployment tracking [Online].
Available: \url{https://secspider.net/stats.html} (Accessed 28.11.2022)}.

In this paper, we leverage the  \ac{DNSSEC} protocol and its operational ecosystem to solve the problem of service authenticity and certificate management in vehicles.
We focus on \ac{SOME/IP} \ac{SD} for automotive service invocation, even though our approach could be transferred to other in-vehicle protocols.
Unlike earlier proposals, which manually pre-provisioned certificates for adding authentication during session establishment~\cite{irrsvssvJR20},  our approach manages security credentials dynamically and is capable of fully functional updates.

We model \ac{SOME/IP} service descriptions as a \ac{DNS} namespace and store  parameters of \ac{SOME/IP} service endpoints in the \ac{DNS}. This allows us to bind certificates to the service names using \ac{DANE}. 
Thereafter, \ac{DNSSEC} ensures authenticity and integrity of the records following a content object security model, which allows for seamless replication of records including caching, as well as credential updates.
In cars, we verify the authenticity of the publisher endpoints with the help of a lightweight challenge-response scheme anchored at the  \ac{DANE} certificates.
We demonstrate the feasibility of our approach by extending the \ac{SOME/IP} reference implementation with access to a local \ac{DNS} resolver for service parameters and certificates during the \ac{SD}.
We compare the performance of our scheme to the reference implementation.

The remainder of this paper is structured as follows.
Section~\ref{sec:background_and_related_work} recaps the \ac{SOME/IP} \ac{SD} and related work on secure discovery of services.
Section~\ref{sec:concept} presents our concept of \ac{DNSSEC}-based \ac{SD} for publisher authenticity.
We evaluate our concept in Section~\ref{sec:eval} and discuss performance results.
Section~\ref{sec:conclusion_and_outlook} concludes with an outlook.

%!TEX root = ../main.tex

\section{In-Car Service Security and Related Work}%
\label{sec:background_and_related_work}
Modern cars have a wide range of heterogeneous services as analyzed in our previous work~\cite{chrmk-qosso-19}.
Among them are \ac{ADAS}, which improve road safety and driving experience, and multimedia applications for infotainment.
Traditionally, \ac{E/E} architectures are rigidly integrated at design time and tightly couple software components to their \acp{ECU}.
As the number of services increases, \acs{E/E} architectures become more complex.
Orchestrating software applications across hardware resources in a dynamic \ac{SOA} allows for a more flexible software architecture~\cite{kakww-dsosa-19}.
This enables shorter innovation cycles, frequent updates, and on-demand installation of services.

Current vehicles are vulnerable to networked attacks via various interfaces including \ac{V2X} communication~\cite{mv-sraas-14,mv-reupv-15}.
In an unprotected network of services, a malicious participant can compromise the communication across the entire network.
This could disrupt the function of safety-related services.

Current automotive systems and protocols were often designed for closed environments~\cite{mrfmsvcJR21} and lack a robust security layer.
The AUTOSAR platform~\cite{rgksoasJR20} advises two major \ac{SOA} solutions for the automotive domain, \ac{SOME/IP} and \ac{DDS}~\cite{omg-dds-15}.
\ac{DDS} supports basic service authenticity~\cite{omg-dds-security-18}, while \ac{SOME/IP}, the most widely deployed protocol in the automotive domain, is tailored to a closely protected automotive environment.
\ac{SOME/IP} \ac{SD} lacks security means~\cite{irrsvssvJR20} including data confidentiality, protection against replay attacks, service authorization and authentication.
We focus on securing \ac{SOME/IP} through service authentication using the established Internet standards \ac{DNSSEC} and \ac{DANE} combined with common authenticity standards.

\subsection{Common Standards for Authentication}
Service authentication mechanisms generate trust by attesting the identity of a service  provider.
The certificate-based X.509 \ac{PKI}~\cite{RFC-5280} uses asymmetric cryptography and a trust anchor.
Certificates contain the public key that proves the identity of an entity, such as a service, and a signed reference to the trusted entity.
A client application requests this certificate and authenticates it using the public key of the trusted instance.
Subsequently, the client verifies the endpoint authenticity via a challenge-response protocol ensuring the entity possesses the private key.

The public \ac{CA} model uses the X.509 \ac{PKI} to attest certificate authenticity of Internet applications.
The \ac{TLS} handshake protocol, for example, verifies the endpoint authenticity of entities.
The main problem with the public \ac{CA} model is that any trusted \ac{CA} can issue a certificate for any domain name~\cite{RFC-6698}. 
Multiple signing \acp{CA} can generate attestation collisions.

\ac{DNSSEC}~\cite{RFC-2535} is a well-established infrastructure to secure the \ac{DNS} against unauthorized modifications of its records.
It adds signature records to the \ac{DNS} that ensure integrity and authenticity of the data stored in plain text records.
Asymmetric cryptography establishes a chain of trust from the root zone to any delegated zone. 
This chain of trust is built along the name hierarchy, though, and remains resistant against attestation collisions, which the Web PKI generates if multiple CAs sign the same resource name.

A robust and mature ecosystem developed during more than 15 years of \ac{DNSSEC} deployment.
This includes not only software and tooling but also a professional practice and thorough analyses of credential maintenance~\cite{otsw-fbtfy-22} including the roll-over of the \ac{DNSSEC} root keys~\cite{mtwhc-rrryr-19}. 
It is noteworthy that \ac{DNSSEC} can also be deployed for private namespace management independent of the global Internet naming hierarchy.

\ac{DANE} enables the binding of certificates to names in the \ac{DNS}.
It uses TLSA records to store certificate data tied to domain names.
The certificate presented by a server must then match against the certificate associated with DNS data to determine the integrity and authenticity of the server.
The security of \ac{DANE} is bound to \ac{DNSSEC} and thus benefits from the inherent chain of trust, which ensures the integrity and authenticity of the TLSA records.

In this work, we modify the \ac{SOME/IP} \ac{SD} to query \ac{DNSSEC}-verified service parameters and connection information stored in the \ac{DNS}.
Further, we bind certificates to services using TLSA records.
We sign all records to obtain signature records and achieve authenticity and integrity of the service parameters and certificates.
The benefit from this approach is a collision-free publisher authenticity that is protected by the well-established DNSSEC infrastructure.

\subsection{\ac{SOME/IP} \acl{SD}}
\ac{SOME/IP} is widely used in automotive networks and is capable of communicating via UDP and TCP transport.
Its design goals include scalability and low resource consumption.
\ac{SOME/IP} uses a publish-subscribe model.
Publishers can notify subscribers about an update or an event that has occurred.
\ac{SOME/IP} \ac{SD} announces and discovers services via multicast.
It also performs the session establishment between publishers and subscribers after a successful subscription.
\ac{SOME/IP} provides no means for service authenticity. 

Figure~\ref{fig:someipsd} shows the \ac{SOME/IP} \ac{SD} sequence for service announcements, discovery,
and subscription.
The \ac{SOME/IP} \ac{SD} uses multicast \codeword{find} and \codeword{offer} messages to request and announce services,
which are described by their ID, instance ID, major and minor version.
An \codeword{offer} entry uses so-called endpoint options to describe how to contact a service.
There are two concepts of discovering a service.
(1)~A publisher periodically updates an \codeword{offer} message, as an \codeword{offer} has a limited lifetime.
(2)~A subscriber requests a service via a \codeword{find} message, whereupon corresponding publisher instances
announce it via an \codeword{offer} message.
After the subscriber receives the \codeword{offer} message, it subscribes to the service via unicast specifying its receiving endpoint description and the desired Eventgroup.
If the publisher can provide this service, it acknowledges the subscription, after which the transmission of the requested data begins.

%!TEX root = ../../main.tex

\newcommand{\participantWidth}{2.9cm}
\newcommand{\participantHeight}{9.5cm}%{10.8cm}
\newcommand{\fullWidth}{8.8cm}

\colorlet{lifeline_color}{black!60}
\begin{figure}
    \centering
\begin{tikzpicture}[remember picture,
    participant/.style={rectangle, draw, line width=.5pt, inner sep=5pt, minimum width=\participantWidth, minimum height=\participantHeight, align=center, fill=CoreGray, rounded corners},
    group/.style={rectangle, draw, line width=.5pt, inner sep=5pt, minimum width=\fullWidth, minimum height=3cm, align=center, fill=LightCoreGray, rounded corners},
    %group_label/.style={chamfered rectangle, draw, chamfered rectangle angle=45, chamfered rectangle corners=south east, line width=.5pt, inner sep=5pt, minimum width=2.2cm, minimum height=0.9cm, align=center, fill=white},
    activation/.style={rectangle, draw, line width=.5pt, minimum width=.3cm, minimum height=1cm, align=center, fill=white}
    ]
    \begin{scope}[on background layer]
        % Publisher
        % \node[participant] (publisher) [fill=gray!30]{};
        \node[rectangle,minimum width=\participantWidth, minimum height=\participantHeight, align=center] (publisher) []{};
        % \node[below=2mm] at (publisher.north) {Publisher};
        % Publisher SD Top
        \node[participant] (publisher_sd_top) [%below=0.9cm, 
        below=0cm, minimum width=1.8cm, minimum height=0.7cm] at (publisher.north) {};
        \node[below=1mm] at (publisher_sd_top.north) {Publisher};

        % Subscriber
        % \fullwidth-2.0*\participantWidth-4.0mm
        % \node[participant] (subscriber) [right=2.6cm of publisher, fill=cyan!30] {};
        \node[rectangle,minimum width=\participantWidth, minimum height=\participantHeight, align=center] (subscriber) [right=2.6cm of publisher]{};
        % \node[below=2mm] at (subscriber.north) {Subscriber};
        % Subscriber SD Top
        \node[participant] (subscriber_sd_top) [%below=0.9cm,
            below=0cm, minimum width=1.8cm, minimum height=0.7cm] at (subscriber.north) {};
        \node[below=1mm] at (subscriber_sd_top.north) {Subscriber};
    \end{scope}

    % Publisher lifeline
    \draw[line width=1.0pt, rounded corners, cap=round, dashed, color=lifeline_color] (publisher_sd_top.south) -- (publisher.south);
    % Subscriber liefeline
    \draw[line width=1.0pt, rounded corners, cap=round, dashed, color=lifeline_color] (subscriber_sd_top.south) -- (subscriber.south);
    
    \begin{scope}[on behind layer]
        % Group consumer triggered
        \node[coordinate] (group_anchor) [] at (publisher_sd_top.south -| publisher.west) {};
        \node[group] (activation_cyclic) [minimum height=1.8cm, below right=3.0mm and -2.0mm of group_anchor] {};
        \node[group] (consumer_triggered) [minimum height=2.8cm, below=4mm of activation_cyclic] {};
        \node[group] (publish_subscribe) [minimum height=2.8cm, below=4mm of consumer_triggered] {};
    \end{scope}

    \begin{scope}[on above layer]
        %Group labels
        %activation/cyclic-refresh
        \newcommand{\slant}{0.2,0.2}
        \newcommand{\nLineCoverOffset}{-0.2mm}
        \newcommand{\pLineCoverOffset}{0.1mm}
        \newcommand{\groupLabelWidth}{4.25}
        \newcommand{\aboveGroupLabel}{-0.9mm}
        \node[coordinate] (g1_start) [] at ($(activation_cyclic.north -| activation_cyclic.west)+(0,-4.0mm)$) {};
        \draw (g1_start) -- ++(\groupLabelWidth,0) -- ++(\slant) -- node(g1_intermediate) [] {} ++(0,0) -- (activation_cyclic.north -| g1_intermediate);
        \draw[line width=2.0pt, color=LightCoreGray] ($(activation_cyclic.north -| publisher)+(0,\nLineCoverOffset)$) -- ($(g1_start -| publisher)+(0,\pLineCoverOffset)$);
        \node[above right=\aboveGroupLabel and 0.0cm of g1_start] {announcment/cyclic refresh};
        %consumer triggered
        \node[coordinate] (g2_start) [] at ($(consumer_triggered.north -| consumer_triggered.west)+(0,-4.0mm)$) {};
        \draw (g2_start) -- ++(\groupLabelWidth,0) -- ++(\slant) -- node(g2_intermediate) [] {} ++(0,0) -- (consumer_triggered.north -| g2_intermediate);
        \draw[line width=2.0pt, color=LightCoreGray] ($(consumer_triggered.north -| publisher)+(0,\nLineCoverOffset)$) -- ($(g2_start -| publisher)+(0,\pLineCoverOffset)$);
        \node[above right=\aboveGroupLabel and 0.0cm of g2_start] {consumer-triggered discovery};
        %publish subscribe
        \node[coordinate] (g3_start) [] at ($(publish_subscribe.north -| publish_subscribe.west)+(0,-4.0mm)$) {};
        \draw (g3_start) -- ++(\groupLabelWidth,0) -- ++(\slant) -- node(g3_intermediate) [] {} ++(0,0) -- (publish_subscribe.north -| g3_intermediate);
        \draw[line width=2.0pt, color=LightCoreGray] ($(publish_subscribe.north -| publisher)+(0,\nLineCoverOffset)$) -- ($(g3_start -| publisher)+(0,\pLineCoverOffset)$);
        \node[above right=\aboveGroupLabel and 0.0cm of g3_start] {publish-subscribe};
    \end{scope}

    \begin{scope}[on glass layer]
        %Calls
        %activation/cyclic-refresh
        \node[coordinate] (offer_start) [] at ($(activation_cyclic.north -| publisher_sd_top)+(0,-1.1cm)$) {};
        \draw[-{Straight Barb[length=1.5mm,angle'=40]}] (offer_start) -- node[pos=0.5, align=center]{\codeword{offer(}service, instance,\\major, minor, endpoint options\codeword{)}} (offer_start -| subscriber_sd_top);
        %consumer triggered
        \node[activation] (find_process) [] at ($(consumer_triggered.north -| publisher_sd_top)+(0,-1.6cm)$) {};
        \draw[{Straight Barb[length=1.5mm,angle'=40]}-] (find_process.73) -- node[pos=0.5, align=center]{\codeword{find(}service, instance,\\major, minor\codeword{)}} (find_process.73 -| subscriber_sd_top);
        \draw[-{Straight Barb[length=1.5mm,angle'=40]}, dashed] (find_process.south) -- node[pos=0.5, align=center]{\codeword{offer(}service, instance, major,\\minor, endpoint options\codeword{)}} (find_process.south -| subscriber_sd_top);
        %publish-subscribe
        \node[activation] (subscriber_process) [] at ($(publish_subscribe.north -| publisher_sd_top)+(0,-1.6cm)$) {};
        \draw[{Straight Barb[length=1.5mm,angle'=40]}-] (subscriber_process.78) -- node[pos=0.5, align=center]{\codeword{subscribe(}service, instance,\\~major, eventgroup, endpoint options\codeword{)}} (subscriber_process.78 -| subscriber_sd_top);
        \draw[-{Straight Barb[length=1.5mm,angle'=40]}, dashed] (subscriber_process.south) -- node[pos=0.5, align=center]{\codeword{subscribeAck(}service,\\instance, major, eventgroup\codeword{)}} (subscriber_process.south -| subscriber_sd_top);
    \end{scope}

\end{tikzpicture}
\caption{Service announcement, discovery, and subscription according to the SOME/IP service discovery protocol.}\label{fig:someipsd}
\end{figure}
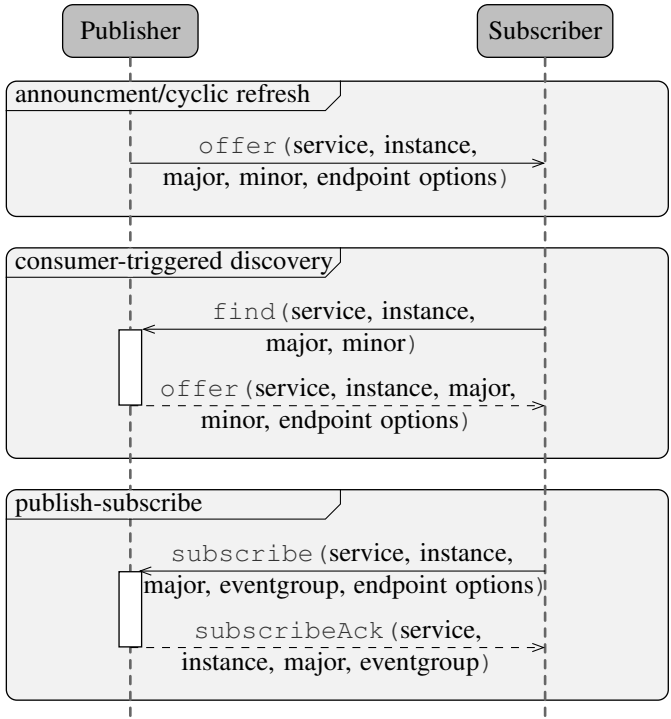

\subsection{In-Vehicle Service Authenticity and Confidentiality}
Secure discovery mechanisms are essential to prevent attackers from infiltrating automotive networks and eavesdropping on in-vehicle communication.
Challenge-response schemes can authenticate nodes to control service access.
Message encryption keeps unauthorized participants from eavesdropping on network communication.
In this work, we focus on publisher authenticity using a challenge-response scheme based on the public credentials obtained from DNSSEC.

Common standards for authentication have been applied to in-vehicle networks.
Challenge-response mechanisms require cryptographic keys that are commonly pre-deployed on the vehicle \acp{ECU} and can be both symmetric secret keys~\cite{kucaeJR22} or asymmetric key pairs~\cite{fizjspsJR17}.
Further, a \ac{PKI} uses a trust anchor to enable the authenticity and integrity of certificates with keys that can be revoked when they are no longer secure~\cite{aahmsapJR21,adpsepJR18}.
In this work, we use asymmetric cryptography for a challenge-response mechanism and \ac{DNSSEC} with its inherent chain of trust to ensure certificate authenticity.

Prior research proposed methods for securing \ac{SOME/IP}, including message encryption and service authentication~\cite{irrsvssvJR20,zlkkassJR21,myzzcascJR22}.
Iorio et al.~\cite{irrsvssvJR20} follow the public \ac{CA} model and a challenge-response scheme using asymmetric cryptography.
Each vehicle has a different trusted root certificate, used to sign the certificates of the \ac{ECU} and the services, creating a simplified chain of trust.
They also bind access control policies to the signed certificates.
We use a similar challenge-response scheme based on asymmetric cryptography to authenticate publishers.
Ma et al.~\cite{myzzcascJR22} use \acp{MAC} against message forgery, and a key management center that derives temporary session keys from encrypted received nonces.
Zelle et al.~\cite{zlkkassJR21} propose two solutions for \ac{SOME/IP} message authentication, one that authenticates services on the \acp{ECU} themselves, and another introducing an authorization server to authenticate messages. 
These solutions, however, require pre-deployed keys and certificates on every \ac{ECU}, which generates the challenge of credential management in practice.
In contrast, we use the \ac{DNS} recursive resolver infrastructure for managing certificate provisioning and \ac{DNSSEC} and \ac{DANE} for ensuring certificate authenticity, reducing the load on the \ac{ECU}.
Our approach only requires signature validation on the \ac{ECU}.

%!TEX root = ../main.tex

\begin{table}[b]
    \caption{SVCB records for one service with symbolic query names and concrete record data.}
    \label{tab:someipsdsvcbrecords}
    \centering
    \normalsize
    \setlength{\tabcolsep}{4pt}        % distance between table columns
    \renewcommand{\arraystretch}{0.85}  % distance between rows
    \begin{tabularx}{\linewidth}{ll} \toprule
        QNAME & RDATA (SVCB) \\
        \midrule
        \_someip.minor.major.instance.id.service.& \multirow{8}{\linewidth}{\parbox{3.0cm}{
                                                                        port=30509\\
                                                                        ipv4hint=10.0.0.5\\
                                                                        protocol=UDP
                                                                        instance=2\\
                                                                        major=1\\
                                                                        minor=2
                                                                        }}\\
        \_someip.major.instance.id.service.&\\
        \textcolor{lightgray}{\_someip.minor.instance.id.service.}&\\
        \_someip.minor.major.id.service.&\\
        \_someip.instance.id.service.&\\
        \_someip.major.id.service.&\\
        \textcolor{lightgray}{\_someip.minor.id.service.}&\\
        \_someip.id.service.&\\
        \bottomrule
    \end{tabularx}
\end{table}

\section{DNSSEC in SOME/IP Service Discovery}\label{sec:concept}
Our approach transforms \ac{SOME/IP} \ac{SD} to utilize the established Internet technologies \ac{DNSSEC} and \ac{DANE} for secure service discovery and authentication.
Therefor we first map \ac{SOME/IP} \ac{SD} data fields to DNS names and record data.
Second, we bind \ac{DANE} certificates to the service names to verify the authenticity.
Our prototype implementation is based on the \textit{vsomeip}~\cite{vsomeip-git} reference implementation, which dictates our architecture and naming.

\subsection{Designing a DNS Namespace for SOME/IP Services}
The main challenges in designing a suitable \ac{DNS} namespace for automotive services are to avoid collisions with existing query names, remain compatible with DNS naming conventions, and simultaneously preserve all \ac{SOME/IP} \ac{SD} query properties.
Four fields specify a service: service ID, instance ID, major version, and minor version.
In a \codeword{find} message, a subscriber must specify at least a service ID, and the other fields can be wildcarded.
For example, not specifying an instance ID results in receiving all running instances of a service.
In total, a service can be requested in $2^3$ ways.

Table~\ref{tab:someipsdsvcbrecords} shows DNS entries for a service based on \ac{SOME/IP} \codeword{find} parameters, using symbolic names for simplicity.
We use \textit{service} as the parent domain, which can be customized, for example, to an OEM or tier-X supplier.
More specifically, adding \textit{tier-x.oem} as the parent domain would enable a hierarchy that passes down the rights to maintain and certify service records in each subdomain.
The four data fields are prepended to the query name in the same order as in the \codeword{find} message service description. 
An unspecified field in a \codeword{find} message corresponds to the absence of that field in the query name.
The arrangement of the four fields is arbitrary, but must be followed consistently as they determine valid query names.
We prepend \textit{\_someip} to each branch, following the semantic scope of the attribute leaf name pattern.

We consider the two gray-marked query names specifying a minor version without a major version impractical and therefore invalid. 
Even though the \ac{SOME/IP} \ac{SD} specification does not object to such queries, the number of valid query names per service is reduced to six.

To prevent ambiguity when wildcarding different fields, we include the symbolic name before the data field value making records uniquely distinguishable. 
As per our namespace design, a service with ID 1, instance ID 2, major version 1, and minor version 2 has this concrete query name:
\smallskip

\centerline{\_someip.minor0x00000002.major0x01.instance0x0002.}
\centerline{id0x0001.service.}
% \medskip

\subsection{Choosing a Record Type for SOME/IP Endpoints}

The IETF specifies various record types for storing data in the \ac{DNS}.
To ensure interoperability between \ac{SOME/IP} \ac{SD} and \ac{DNSSEC}, we need a DNS record that can contain all information originally provided in \codeword{offer} messages.
This includes the service description (Figure~\ref{fig:find-offer-entry}) and additional endpoint options (Figure~\ref{fig:endpointoption}), specifying how to connect to a service using an IP address, L4-protocol, and port number.

%!TEX root = ../../main.tex

\newcommand{\bytegap}{4.2mm}
\newcommand{\datagap}{9.5mm}
\begin{figure} 
    \centering
    \begin{tikzpicture}[remember picture,
    full_width/.style={rectangle, draw, line width=.5pt, inner sep=5pt, minimum width=8.8cm, minimum height=1.55cm, align=center, fill=white, rounded corners},
    rounded_rectangle/.style={rectangle, draw, line width=.5pt, inner sep=3pt, align=center, fill=CoreGray, rounded corners, minimum width=2.21cm, minimum height=1.55cm},
    rounded_rectangle_data/.style={rectangle, line width=.5pt, inner sep=3pt, align=center, fill=CoreMiddleGray, rounded corners=2.4mm, minimum width=1.8cm, minimum height=0.5cm}
    ]

    \begin{scope}[on background layer]
        \node[full_width, draw=none] (find_outline) [] {};
    \end{scope}

    %main layer
    % service id
    \node[rounded_rectangle] (service_id_outline) [right=-0.5pt of find_outline.west, align=center] {};
    \node(service_id_text) [below=0.0cm of service_id_outline.north, align=center] {service ID};
    \node(service_id_byte) [below=\bytegap of service_id_outline.north, align=center] {\footnotesize{(2 Byte)}};
    \node[rounded_rectangle_data] (service_id_data) [below=\datagap of service_id_outline.north, align=center] {0x0001};

    % instance id
    \node[rounded_rectangle] (instance_id_outline) [right=-0.5pt of service_id_outline, align=center] {};
    \node(instance_id_text) [below=0.0cm of instance_id_outline.north, align=center] {instance ID};
    \node(instance_id_byte) [below=\bytegap of instance_id_outline.north, align=center] {\footnotesize{(2 Byte)}};
    \node[rounded_rectangle_data] (instance_id_data) [below=\datagap of instance_id_outline.north, align=center] {0x0002};

    % major version
    \node[rounded_rectangle] (major_id_outline) [right=-0.5pt of instance_id_outline, align=center] {};
    \node(major_id_text) [below=0.0cm of major_id_outline.north, align=center] {major version};
    \node(major_id_byte) [below=\bytegap of major_id_outline.north, align=center] {\footnotesize{(1 Byte)}};
    \node[rounded_rectangle_data] (major_id_data) [below=\datagap of major_id_outline.north, align=center] {0x01};

    % minor version
    \node[rounded_rectangle] (minor_id_outline) [right=-0.5pt of major_id_outline, align=center] {};
    \node(minor_id_text) [below=0.0cm of minor_id_outline.north, align=center] {minor version};
    \node(minor_id_byte) [below=\bytegap of minor_id_outline.north, align=center] {\footnotesize{(4 Byte)}};
    \node[rounded_rectangle_data] (minor_id_data) [below=\datagap of minor_id_outline.north, align=center] {0x00000002};

    \begin{scope}[on behind layer]
    \end{scope}

    \begin{scope}[on above layer]
    \end{scope}

    \begin{scope}[on glass layer]
    \end{scope}

\end{tikzpicture}
\caption{
    Service description in \codeword{find} and \codeword{offer} messages.
    %  for a service with ID 1, instance ID 2, major version 1 and minor version 2
    }
\label{fig:find-offer-entry}
\end{figure}
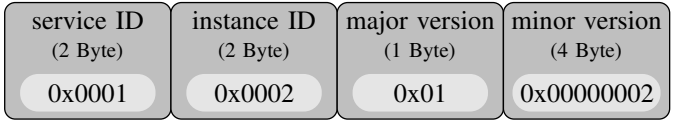
%!TEX root = ../../main.tex

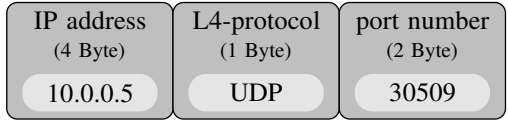
\begin{figure} 
    \centering
    \begin{tikzpicture}[remember picture,
    full_width/.style={rectangle, draw, line width=.5pt, inner sep=5pt, minimum width=6.6cm, minimum height=1.55cm, align=center, fill=white, rounded corners},
    rounded_rectangle/.style={rectangle, draw, line width=.5pt, inner sep=3pt, align=center, fill=CoreGray, rounded corners, minimum width=2.21cm, minimum height=1.55cm},
    rounded_rectangle_data/.style={rectangle, line width=.5pt, inner sep=3pt, align=center, fill=CoreMiddleGray, rounded corners=2.4mm, minimum width=1.8cm, minimum height=0.5cm}
    ]

    \begin{scope}[on background layer]
        \node[full_width, draw=none] (endpoint_option_outline) [] {};
    \end{scope}

    %main layer
    % ip address
    \node[rounded_rectangle] (ip_address_outline) [right=-0.5pt of endpoint_option_outline.west, align=center] {};
    \node(ip_address_text) [below=0.0cm of ip_address_outline.north, align=center] {IP address};
    \node(ip_address_byte) [below=\bytegap of ip_address_outline.north, align=center] {\footnotesize{(4 Byte)}};
    \node[rounded_rectangle_data] (ip_address_data) [below=\datagap of ip_address_outline.north, align=center] {10.0.0.5};

    % transport protocol
    \node[rounded_rectangle] (transport_protocol_outline) [right=-0.5pt of ip_address_outline, align=center] {};
    \node(transport_protocol_text) [below=0.0cm of transport_protocol_outline.north, align=center] {L4-protocol};
    \node(transport_protocol_byte) [below=\bytegap of transport_protocol_outline.north, align=center] {\footnotesize{(1 Byte)}};
    \node[rounded_rectangle_data] (transport_protocol_data) [below=\datagap of transport_protocol_outline.north, align=center] {UDP};

    % port number
    \node[rounded_rectangle] (port_number_outline) [right=-0.5pt of transport_protocol_outline, align=center] {};
    \node(port_number_text) [below=0.0cm of port_number_outline.north, align=center] {port number};
    \node(port_number_byte) [below=\bytegap of port_number_outline.north, align=center] {\footnotesize{(2 Byte)}};
    \node[rounded_rectangle_data] (port_number_data) [below=\datagap of port_number_outline.north, align=center] {30509};

    \begin{scope}[on behind layer]
    \end{scope}

    \begin{scope}[on above layer]
    \end{scope}

    \begin{scope}[on glass layer]
    \end{scope}

\end{tikzpicture}
\caption{
    Endpoint details in \codeword{offer} and \codeword{subscribe} messages.
    %  for a service with IP address 10.0.0.5, layer 4 protocol UDP and port number 30509
    }
\label{fig:endpointoption}
\end{figure}

The main \ac{DNS} service record candidates are SRV and SVCB.
The SRV record~\cite{RFC-2782} specifies a service endpoint location, including a transport port and domain name data fields.
SRV record names should follow attribute leaf naming~\cite{RFC-8552}, using underscored names prepended to the parent domain for semantic classification of services, for example, including the transport protocol in the name (e.g., \mbox{\_ldap.\_tcp.example.com}). 
This does not comply with \ac{SOME/IP} \ac{SD} as the transport protocol is not specified in \codeword{find} messages.
Next, the SRV record does not provide a data field for the IP address of a service, but holds a domain name referencing an address record instead, which requires additional queries as a detour to obtain an address and available transport protocols.
The SVCB record~\cite{sbnsbpJR22} stores general purpose service bindings and is still an active IETF Internet-Draft in the converging phase to become a standard.
SVCB data includes fields for port number, IP address and 255 other fields for private use to store service parameters, making it our preferred option.

Table~\ref{tab:someipsdsvcbrecords} shows the SVCB record data uniform for all query names of a single service, with \codeword{offer} message fields and endpoint options mapped accordingly.
The record data refers to an UDP-accessible service at port 30509 and IP address 10.0.0.5 stored as ipv4hint, with instance ID 2, major version 1, and minor version 2 specified in case wildcards were used in the query name.
The instance ID, major version, minor version and layer 4 protocol are each mapped to one of the 255 fields for private use.

\subsection{Integrating the SOME/IP SD for using DNSSEC}
\ac{DNSSEC} ensures records are unchanged and correct when the subscriber receives them.
This is already an advantage over the \ac{SOME/IP} \ac{SD}, where anyone can send conflicting offers~\cite{zlkkassJR21}.
Our approach showcases the adaptation of a \ac{SOME/IP} stack for \ac{DNSSEC}-based service discovery using the open source reference implementation \textit{vsomeip}~\cite{vsomeip-git}.

\begin{figure} 
    \centering
    \begin{tikzpicture}[remember picture,
    full_width/.style={rectangle, draw, line width=.5pt, inner sep=5pt, minimum width=8.8cm, minimum height=3cm, align=center, fill=white, rounded corners},
    rounded_rectangle/.style={rectangle, line width=.5pt, inner sep=3pt, align=center, fill=white, rounded corners}
    ]

    \begin{scope}[on background layer]
        % Service Discovery
        % Server
        \node[full_width] (server) [minimum height=4.44cm, minimum width=3.7cm, fill=CoreGray] {};
        \node[coordinate] (server_start) [below left=0.5cm and 1.85cm of server.north] {};
        \draw (server_start) -- ++(1.05,0) -- ++(0.2,0.2) -- ++(0,0.3);
        \node[above right=0.1mm and 0.0mm of server_start] {\textbf{Server}};
        \node[rounded_rectangle] (pub_app) [below=6.0mm of server.north, minimum width=3.5cm, fill=CoreMiddleGray] {Publisher Application};
        \node[rounded_rectangle] (routing_manager_pub) [below=1.0mm of pub_app, minimum width=3.5cm, fill=CoreMiddleGray] {Routing Manager};
        \node[rounded_rectangle] (sd_pub) [below=1.0mm of routing_manager_pub, minimum height=2.45cm, minimum width=3.5cm, fill=CoreMiddleGray] {};
        \node(sd_pub_text) [below left=0.0mm and -1.1cm of sd_pub.north] {Service Discovery};
        \node[rounded_rectangle] (offer_find_pub) [below left=5.0mm and 0.5mm of sd_pub.north, align=center, minimum width=1.6cm, draw=none, text=gray] {Offer/\\Find};
        \node[rounded_rectangle] (pub_sub_pub) [right=1.0mm of offer_find_pub, minimum width=1.6cm, minimum height=0.87cm] {PubSub};
        \node[rounded_rectangle] (dnssec_pub) [below=1.0mm of offer_find_pub, minimum width=1.6cm, minimum height=0.87cm, line width=1pt, draw=CoreRed] {DNSSEC};

        % Client
        \node[full_width] (client) [right=1.4cm of server, minimum height=4.44cm, minimum width=3.7cm, fill=CoreGray] {};
        \node[coordinate] (client_start) [below left=0.5cm and 1.85cm of client.north] {};
        \draw (client_start) -- ++(1.05,0) -- ++(0.2,0.2) -- ++(0,0.3);
        \node[above right=0.1mm and 0.0mm of client_start] {\textbf{Client}};
        \node[rounded_rectangle] (sub_app) [below=6.0mm of client.north, fill=CoreMiddleGray] {Subscriber Application};
        \node[rounded_rectangle] (routing_manager_sub) [below=1.0mm of sub_app, minimum width=3.5cm, fill=CoreMiddleGray] {Routing Manager};
        \node[rounded_rectangle, fill=CoreMiddleGray] (sd_sub) [below=1.0mm of routing_manager_sub, minimum height=2.45cm, minimum width=3.5cm] {};
        \node (sd_sub_text) [below left=0.0mm and -1.1cm of sd_sub.north] {Service Discovery};
        \node[rounded_rectangle] (pub_sub_sub) [below left=5.0mm and 0.5mm of sd_sub.north, minimum width=1.6cm, minimum height=0.87cm] {PubSub};
        \node[rounded_rectangle] (offer_find_sub) [right=1.0mm of pub_sub_sub, align=center, minimum width=1.6cm, draw=none, text=gray] {Offer/\\Find};
        \node[rounded_rectangle] (dnssec_sub) [below=1.0mm of offer_find_sub, minimum width=1.6cm, minimum height=0.87cm, line width=1pt, draw=CoreRed] {DNSSEC};

        % DNSSEC Recursive Resolver
        \node[full_width] (dns_resolver) [below right=2.0mm and 0.0mm of server.south, minimum height=11.5mm, minimum width=4.4cm, fill=CoreGray] {};
        \node[coordinate] (dns_resolver_start) [below left=0.5cm and 2.2cm of dns_resolver.north] {};
        \draw (dns_resolver_start) -- ++(2.9,0) -- ++(0.2,0.2) -- ++(0,0.3);
        \node[above right=0.0mm and 0.0cm of dns_resolver_start.south] {\textbf{DNSSEC Resolver}};
        \node[rounded_rectangle] (svcb_records) [below=6.0mm of dns_resolver.north, minimum width=4.2cm, fill=CoreMiddleGray] {Service and DANE Records};

        % Arrows and Lines
        \draw[latex-latex] (pub_sub_pub.south) -- ++(0,-0.5) -- node[pos=0.5, align=center] {2. Sub\\SubAck} ($(pub_sub_sub.south)+(0,-0.5)$) -- (pub_sub_sub.south);
        \draw[-latex] (pub_app.east) -- node(pub_data_text)[pos=0.5, align=center] {3.Publish\\Data} (sub_app.west);
        \draw[-latex, line width=1pt, draw=CoreRed, text=CoreRed] (dnssec_sub.south) -- (dnssec_sub.south |- dns_resolver.east)-- node(dnssec_intermediate) [above, align=center] {1. resolve} (dns_resolver.east);
        \draw[-, draw=darkgray] (offer_find_pub.north -| offer_find_pub.west) -- (offer_find_pub.south -| offer_find_pub.east);
        \draw[-, draw=darkgray] (offer_find_pub.south -| offer_find_pub.west) -- (offer_find_pub.north -| offer_find_pub.east);
        \draw[-, draw=darkgray] (offer_find_sub.north -| offer_find_sub.west) -- (offer_find_sub.south -| offer_find_sub.east);
        \draw[-, draw=darkgray] (offer_find_sub.south -| offer_find_sub.west) -- (offer_find_sub.north -| offer_find_sub.east);
    \end{scope}

    %main layer
    
    \begin{scope}[on behind layer]
    \end{scope}

    \begin{scope}[on above layer]
    \end{scope}

    \begin{scope}[on glass layer]
    \end{scope}

\end{tikzpicture}
\caption{\acs{SOME/IP} \acs{SD} modification for using \acs{DNSSEC}.}
\label{fig:someipsddnsmod}
\end{figure}
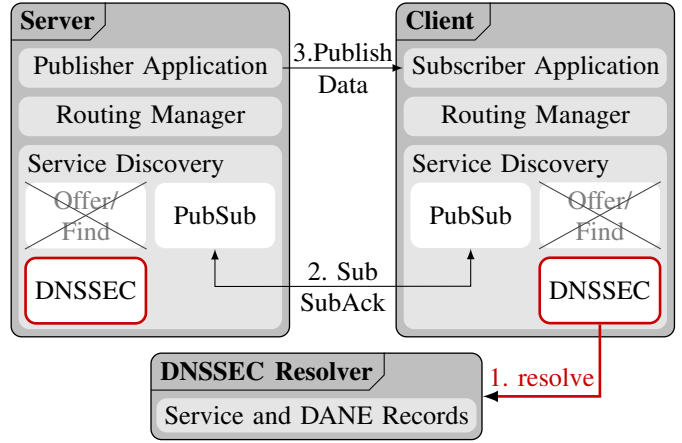

Figure~\ref{fig:someipsddnsmod} depicts the conceptual architecture inherited from the \textit{vsomeip} stack, comprising an application, routing manager, and service discovery used by both the client and server.
The routing manager handles the local transport-specific endpoints for the applications and forwards messages between them. 

Our modifications to enable \ac{DNSSEC} during \ac{SOME/IP} \ac{SD} are also illustrated in Figure~\ref{fig:someipsddnsmod}.
Instead of the original \codeword{offer}/\codeword{find} procedure, the client retrieves the publisher endpoint description via a \ac{DNSSEC} resolver.
With that, publisher services no longer announce themselves, and we gain secure service discovery through the implicit trust established by \ac{DNSSEC} records.
Clients can subscribe to the service and receive published data after obtaining the service record.

%!TEX root = ../../main.tex

% \definecolor{DNScolor}{HTML}{c80000}
% \definecolor{dark_gray}{HTML}{bfbfbf}
% \definecolor{middle_gray}{HTML}{e5e5e5}

\subsection{Ensuring Publisher Authenticity with DANE}
\ac{DNSSEC} verifies subscription parameters used to access the endpoint.
Nevertheless, attackers can still mimic this endpoint, for example, using IP spoofing.
\ac{DANE} validates publisher authenticity to ensure that the subscriber connects to the correct publisher.

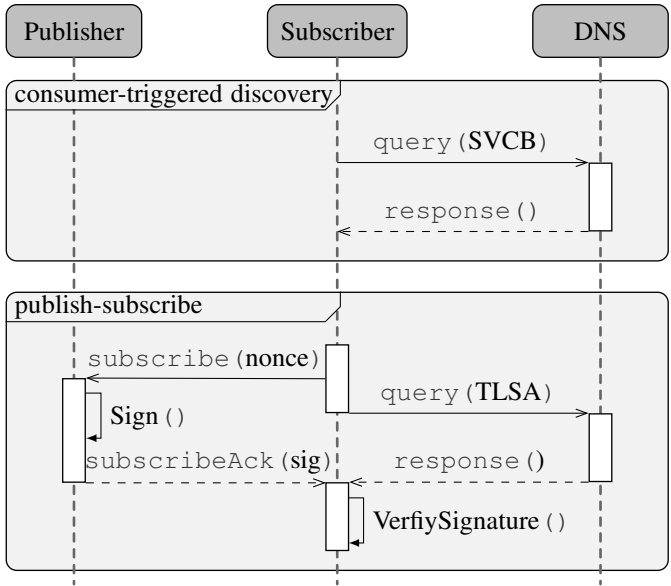
\begin{figure}
    \centering
\begin{tikzpicture}[remember picture,
    participant/.style={rectangle, draw, line width=.5pt, inner sep=5pt, minimum width=0.0cm, minimum height=0.0cm, align=center, fill=CoreGray, rounded corners},
    group/.style={rectangle, draw, line width=.5pt, inner sep=5pt, minimum width=\fullWidth, minimum height=3cm, align=center, rounded corners, fill=LightCoreGray},
    activation/.style={rectangle, draw, line width=.5pt, minimum width=.3cm, minimum height=0.9cm, align=center, fill=white}
    ]
    \begin{scope}[on background layer]
        % Publisher
        \node[participant, minimum width=1.8cm, minimum height=0.7cm] (publisher) {Publisher};
        % Subscriber
        \node[participant, minimum width=1.8cm, minimum height=0.7cm] (subscriber) [right=1.655cm of publisher] {Subscriber};
        % DNS
        \node[participant, minimum width=1.8cm, minimum height=0.7cm] (dns) [right=1.655cm of subscriber] {DNS};
    \end{scope}

    % Publisher lifeline
    \draw[line width=1.0pt, rounded corners, cap=round, dashed, color=lifeline_color] (publisher.south) -- ++(0,-7.0);
    % Subscriber liefeline
    \draw[line width=1.0pt, rounded corners, cap=round, dashed, color=lifeline_color] (subscriber.south) -- ++(0,-7.0);
    % DNS liefeline
    \draw[line width=1.0pt, rounded corners, cap=round, dashed, color=lifeline_color] (dns.south) -- ++(0,-7.0);
    
    \begin{scope}[on behind layer]
        \node[coordinate] (group_anchor) [] at (publisher.west) {};
        \node[group] (consumer_triggered) [minimum height=2.4cm, below right=6.5mm and 0.0mm of group_anchor] {};
        \node[group] (publish_subscribe) [minimum height=3.7cm, below=4mm of consumer_triggered] {};
    \end{scope}

    \begin{scope}[on above layer]
        \newcommand{\selfMessageOffset}{-0.2cm}
        \newcommand{\selfMessageSpacing}{0.2cm}
        \newcommand{\selfMessageDuration}{-0.8cm}
        %Calls
        %SVCB request
        \node[coordinate] (request_start) [] at ($(consumer_triggered.north -| subscriber)+(0,-1.1cm)$) {};
        \node[activation] (request_process) [below] at (request_start -| dns) {};
        \node[coordinate] (request_process_nw) at (request_process.north west) {};
        \draw[-{Straight Barb[length=1.5mm,angle'=40]}] (request_process_nw -| request_start) -- node[pos=0.5, align=center, above] {\codeword{query(}SVCB\codeword{)}} (request_process_nw);
        %SVCB response
        \node[coordinate] (request_process_sw) at (request_process.south west) {};
        \node[coordinate] (response_process_ne) [] at (request_process_sw -| subscriber) {};
        \draw[-{Straight Barb[length=1.5mm,angle'=40]}, dashed] (request_process_sw) -- node[pos=0.5, align=center, above]{\codeword{response()}} (response_process_ne);
        
        %TLSA request
        \node[activation] (response_process) [below] at ($(publish_subscribe.north -| subscriber)+(0,-0.7cm)$) {};
        \node[coordinate] (response_process_se) at (response_process.south east) {};
        \node[activation] (request_tlsa_process) [below] at (response_process_se -| dns) {};
        \draw[-{Straight Barb[length=1.5mm,angle'=40]}] (response_process_se) -- node[pos=0.5, align=center, above] {\codeword{query(}TLSA\codeword{)}} (request_tlsa_process.north west);
        %\codeword{subscribe}
        \node[coordinate] (subscribe_start) at (response_process.west) {};
        \node[coordinate] (subscribe_process_nw) at (subscribe_start -| publisher) {};
        \node[coordinate] (subscribe_process_se) at (request_tlsa_process.south -| publisher) {};
        \node[activation, inner sep=0pt, fit=(subscribe_process_nw)(subscribe_process_se)] (subscribe_process) {};
        % \node[activation, minimum height=1.5cm] (subscribe_process) [below] at (subscribe_process_nw) {};
        \node[coordinate] (subscribe_process_ne) at (subscribe_process.north east) {};
        \draw[-{Straight Barb[length=1.5mm,angle'=40]}] (subscribe_start) -- node[pos=0.5, align=center, above] {\codeword{subscribe(}nonce\codeword{)}} (subscribe_process_ne);
        %Signing
        \node[coordinate] (pub_sign_upper_right_arrow) at ($(subscribe_process.north east)+(\selfMessageSpacing,\selfMessageOffset)$) {};
        \node[coordinate] (pub_sign_lower_right_arrow) at ($(subscribe_process.north east)+(\selfMessageSpacing, \selfMessageDuration)$) {};
        \draw[-latex] ($(subscribe_process.north east)+(0.0cm,\selfMessageOffset)$) -- (pub_sign_upper_right_arrow) -- (pub_sign_lower_right_arrow) -- ($(subscribe_process.north east)+(0.0cm,\selfMessageDuration)$);
        \node[below right=0.5mm and 0mm of pub_sign_upper_right_arrow, align=left]{Sign\codeword{()}};
        %\codeword{subscribe} ACK and TLSA response
        \node[activation] (sub_tlsa_process) [below] at (request_tlsa_process.south -| subscriber) {};
        \draw[-{Straight Barb[length=1.5mm,angle'=40]}, dashed] (request_tlsa_process.south west) -- node[pos=0.5, align=center, above]{\codeword{response(})} (sub_tlsa_process.north east);
        \draw[-{Straight Barb[length=1.5mm,angle'=40]}, dashed] (subscribe_process.south east) -- node[pos=0.5, align=center, above]{~\codeword{subscribeAck(}sig\codeword{)}} ($(subscribe_process.south -| subscriber)+(-1.5mm,0.0mm)$);
        %TLSA validation
        \node[coordinate] (sub_tlsa_upper_right_arrow) at ($(subscribe_process.south -| subscriber)+(1.5mm,0.0mm)+(\selfMessageSpacing,\selfMessageOffset)$) {};
        \node[coordinate] (sub_tlsa_lower_right_arrow) at ($(subscribe_process.south -| subscriber)+(1.5mm,0.0mm)+(\selfMessageSpacing, \selfMessageDuration)$) {};
        \draw[-latex] ($(subscribe_process.south -| subscriber)+(1.5mm,0.0mm)+(0.0cm,\selfMessageOffset)$) -- (sub_tlsa_upper_right_arrow) -- (sub_tlsa_lower_right_arrow) -- ($(subscribe_process.south -| subscriber)+(1.5mm,0.0mm)+(0.0cm,\selfMessageDuration)$);
        \node[below right=0.5mm and 0mm of sub_tlsa_upper_right_arrow, align=left]{VerfiySignature\codeword{()}};
    \end{scope}

    \begin{scope}[on glass layer]
        \newcommand{\slant}{0.2,0.2}
        \newcommand{\nLineCoverOffset}{-0.2mm}
        \newcommand{\pLineCoverOffsetPub}{0.1mm}
        \newcommand{\pLineCoverOffsetSub}{1.5mm}
        \newcommand{\groupLabelWidth}{4.25}
        \newcommand{\aboveGroupLabel}{-0.9mm}
        %consumer triggered
        \node[coordinate] (g1_start) [] at ($(consumer_triggered.north -| consumer_triggered.west)+(0,-4.0mm)$) {};
        \draw[line width=2.0pt, color=LightCoreGray] ($(consumer_triggered.north -| publisher)+(0,\nLineCoverOffset)$) -- ($(g1_start -| publisher)+(0,\pLineCoverOffsetPub)$);
        \draw[line width=2.0pt, color=LightCoreGray] ($(consumer_triggered.north -| subscriber)+(0,\nLineCoverOffset)$) -- ($(g1_start -| subscriber)+(0,\pLineCoverOffsetSub)$);
        \draw(g1_start) -- ++(\groupLabelWidth,0) -- ++(\slant) -- node(g1_intermediate) [] {} ++(0,0) -- (consumer_triggered.north -| g1_intermediate);
        \node[above right=\aboveGroupLabel and 0.0cm of g1_start] {consumer-triggered discovery};

        %pubsub
        \node[coordinate] (g2_start) [] at ($(publish_subscribe.north -| publish_subscribe.west)+(0,-4.0mm)$) {};
        \draw[line width=2.0pt, color=LightCoreGray] ($(publish_subscribe.north -| publisher)+(0,\nLineCoverOffset)$) -- ($(g2_start -| publisher)+(0,\pLineCoverOffsetPub)$);
        \draw[line width=2.0pt, color=LightCoreGray] ($(publish_subscribe.north -| subscriber)+(0,\nLineCoverOffset)$) -- ($(g2_start -| subscriber)+(0,\pLineCoverOffsetSub)$);
        \draw(g2_start) -- ++(\groupLabelWidth,0) -- ++(\slant) -- node(g2_intermediate) [] {} ++(0,0) -- (publish_subscribe.north -| g2_intermediate);
        \node[above right=\aboveGroupLabel and 0.0cm of g2_start] {publish-subscribe};

    \end{scope}

\end{tikzpicture}
\caption{Augmented SOME/IP SD with DNSSEC and DANE for secure publisher service discovery and authentication.}\label{fig:someipsdaugdns}
\end{figure}

Figure~\ref{fig:someipsdaugdns} shows our secure service discovery and invocation process.
After the subscriber has resolved the service endpoint through a DNS \codeword{query}, it subscribes to the service with the information from the SVCB record.
At the same time, the subscriber queries the DNS for the DANE TLSA record containing the public certificate of the service, which is again protected with \ac{DNSSEC}.
With this the subscriber can validate the signature of the publisher.

We employ a challenge-response scheme to ensure that the publisher endpoint is authentic and indeed the owner of the corresponding private key.
With the \codeword{subscribe} message, the subscriber sends a random 32-bit nonce in a \ac{SOME/IP} configuration option as a challenge to the publisher. 
The publisher signs the challenge with its private key and sends the subscription acknowledgement with the signed random nonce back to the subscriber, also utilizing a configuration option.
The subscriber validates the signature with the public certificate of the publisher, providing assurance of authenticity.
With a future extension, the subscriber and publisher could agree on a session key during the challenge-response process to enable message encryption for confidentiality.

%!TEX root = ../../main.tex

\subsection{Operating DNS-based Automotive Service Discovery}
In operation, we foresee that a car has a local \ac{DNSSEC} recursive resolver that caches verified records as soon as the car has Internet connectivity.
Each time a record is retrieved, the \ac{DNSSEC} recursive resolver ensures the chain of trust before caching it, eliminating the need for \ac{DNSSEC} validation during \ac{SD}.
This ensures that the service discovery is still operational when the vehicle is disconnected from the Internet.
Cached records are refreshed before they expire, and it shall be part of future experimentally driven research to determine appropriate cache lifetimes in real deployments.
In this way, our approach exploits the benefits of a well-established standard infrastructure for obtaining data integrity, authenticity, and a robust procedure for certificate management.

%!TEX root = ../main.tex

\section{Discovery Capabilities and Performance}
\label{sec:eval}
We evaluate the performance of the proposed solution compared to the unchanged SOME/IP SD protocol.
Therefore, we first compare the service discovery capabilities showing differences in communication schemes and security mechanisms.
Then, we evaluate the performance of our prototype implementation in terms of discovery and subscription latency.

\begin{table*}
    \caption{Feature comparison between the SOME/IP SD protocol and the proposed approach based on DNSSEC and DANE.}
    \label{tab:comparison}
    \centering
    \normalsize
    \setlength{\tabcolsep}{7pt}        % distance between table columns
    \begin{tabularx}{\linewidth}{p{4.2cm} L L} \toprule
        \textbf{Feature} & \textbf{SOME/IP SD} (and related work) & \textbf{SD w/ DNSSEC and DANE}~(our~approach)\\
        \midrule
        Standard commission & 
        AUTOSAR~\cite{aspsJR21,assdJR21} & 
        IETF~\cite{RFC-2535,RFC-6698,sbnsbpJR22}\\ 

        \addlinespace[7pt]

        Introduction and deployment &
        Basic support in AUTOSAR since Nov. 2016, deployment in production vehicles just starting &
        DNSSEC first standardized in 1997, over 15 years of global deployment and operational experience\\
        
        \addlinespace[7pt]

        Target environment &
        Developed for local in-vehicle network &
        Hardened for global Internet deployment\\

        \addlinespace[7pt]

        Service discovery scheme & 
        Multicast \codeword{find}/\codeword{offer} messages &
        Unicast DNS \codeword{query}/\codeword{response} \\

        \addlinespace[7pt]

        Endpoint detail distribution &
        Provided initiated \codeword{offer} messages with the service runtime location &
        Consumer requested DNSSEC-signed SVCB records with pre-defined endpoint information \\

        \addlinespace[7pt]

        Authentication scheme &
        None by default, challenge-response~\cite{irrsvssvJR20,zlkkassJR21}, central authorization server~\cite{zlkkassJR21,myzzcascJR22} &
        Challenge-response during subscription \\ 

        \addlinespace[7pt]

        Certificate distribution &
        Pre-deployed public certificates on all participating nodes~\cite{irrsvssvJR20,zlkkassJR21,myzzcascJR22} &
        Consumer requested public certificates from DNSSEC-signed TLSA records \\

        \addlinespace[7pt]

        Certificate update procedure &
        No automated mechanism, requires simultaneous update of all cients and servers in workshop or OTA~\cite{irrsvssvJR20,zlkkassJR21} &
        Established mechanism for certificate management, including update and revocation\\

        \bottomrule
    \end{tabularx}
\end{table*}

\subsection{Key Features and Implications}\label{sec:discussion}

Table~\ref{tab:comparison} summarizes differences in key features between the proposed approach using DNSSEC and DANE, and the \ac{SOME/IP} \ac{SD} protocol.
\ac{SOME/IP} was initially released in 2016 as a module in the AUTOSAR platform and targets local in-vehicle networks.
\ac{DNSSEC} and \ac{DANE} are defined in RFCs by the IETF.
Our approach with \ac{DNSSEC} and \ac{DANE} leverages this technology with over 15 years of global deployment and operational experience on the Internet.
With this we gain the benefits of a tried, resilient and security hardened infrastructure.

The \ac{SOME/IP} \ac{SD} uses group communication, whereas the \ac{DNS} uses unicasts.
For \ac{DNS}-based discovery, this implies that multiple clients of the same server must all query the \ac{DNS} resolver separately, while with \ac{SOME/IP} \ac{SD}, a publisher can inform subscribers with a single multicast \codeword{offer}, reducing the network load.
An evaluation in a realistic automotive setup with a large number of services would show whether our approach introduces significant performance penalties, but we leave that open for future work.

Endpoint discovery in the \ac{SOME/IP} \ac{SD} uses \codeword{offer} messages initiated by the publisher, while with \ac{DNS}, the consumer queries the resolver directly.
This means that publishers can provide endpoint information during runtime with \ac{SOME/IP} \ac{SD}, but the \ac{DNS} resolver is not aware of the service runtime location.
This requires predefined IP addresses and ports for all service instances in \ac{DNS} records, which should be the same for all vehicles. 
However, this is not a problem within a local network where the IP addresses can be freely selected.
In turn, the \ac{DNS} records can be verified along the \ac{DNSSEC} trust chain, which is not possible for the endpoint information provided by the publisher.
This prevents malicious services to \codeword{offer} false endpoint information, for which there is no protection with \ac{SOME/IP} \ac{SD}.

There are no service authentication means in \ac{SOME/IP} \ac{SD} by default.
Previous work proposed using the public \ac{CA} model~\cite{irrsvssvJR20,zlkkassJR21} or a custom key management center~\cite{myzzcascJR22} to ensure the authenticity of public certificates pre-deployed on each node in an additional challenge-response handshake.
Our approach exploits the \ac{DNSSEC} and \ac{DANE} mechanisms as an established standard to ensure implicit certificate and service information authenticity through the \ac{DNSSEC} trust chain.
We integrate a challenge-response mechanism similar to the \ac{TLS} handshake into the \ac{SOME/IP} subscription process ensuring endpoint authenticity.
With this, we have to perform additional lookups in the \ac{DNS} to retrieve the certificate and the service information from an in-vehicle \ac{DNSSEC} resolver. 
In our benchmark, we evaluate the discovery latency showing that the certificate lookups do not introduce a significant overhead. 

A vital advantage in using \ac{DNSSEC} and \ac{DANE} mechanisms are the well-established certificate management procedures.
Easy online updates are performed by adding new records to the \ac{DNS} for the new version of the app and new keys on the publisher node can be deployed with a new version of the application.
This ensures that older versions can still use the old certificates while the already updated applications can use the new certificates.
In contrast, pre-deployed keys must be updated on every client and server node.
In case of certificate changes that also affect the private keys, the private key in question must be updated, both for pre-deployed certificates with SOME/IP and for certificates in TLSA records.

Previous work investigated attack vectors on the \ac{SOME/IP} {SD} protocol, such as \ac{MITM} attacks exploiting \codeword{offer} messages~\cite{zlkkassJR21}.
Our consumer-triggered discovery approach with known IP addresses from the \ac{DNS} is not affected by the identified attacks, although an attacker using ARP and IP spoofing could act as a malicious proxy altering the messages when no encryption or signature is used.
We assume that the local \ac{DNS} resolver is only accessible from the in-vehicle network and therefore not easily manipulated, which also prevents potential \ac{DoS} attacks from the outside.
Even if the uplink of the \ac{DNS} is jammed, the in-vehicle network communication can still operate.
However, if the vehicle is operated for an extended period of time without an Internet connection, the authenticity of the service cannot be guaranteed, as the certificates could expire or be revoked in the meantime.
\acp{SPOF} can be faced by replicating the local \ac{DNS} resolver as well as the external \ac{DNSSEC} server.

\begin{figure*}[t]
    \centering
    \begin{minipage}[t]{.485\textwidth}
        %!TEX root = ../../main.tex
% \begin{figure}
%     \centering
    \begin{tikzpicture}
        \begin{axis}[
            width=\linewidth,height=.8\linewidth,
            boxplot/draw direction=y,
            xtick = {1,2,3,4},
            xticklabels = {SOME/IP SD, SOME/IP SD\\ w/ AUTH, DNSSEC, DNSSEC \\w/ DANE},
            xticklabel style = {anchor=north east,align=center, font=\small, rotate=35},
            ylabel={Discovery latency},
            y unit={\milli\second},
            enlarge y limits,
            ymajorgrids,
            % xtick style = {draw=none}, % Hide tick line
            % extra y ticks={5},
            % extra y tick labels={\textcolor{CoreBlue}{5}},
            boxplotcolor/.style={color=#1,fill=#1,mark options={color=#1,fill=#1,draw=black}},
            ]
        \addplot+ [boxplotcolor=CoreRed, boxplot, draw=black] table [col sep={comma},y={DISCOVERY_LATENCY}] {data/export/vanilla_no_auth.csv};
        \addplot+ [boxplotcolor=CoreYellow, boxplot, draw=black] table [col sep={comma},y={DISCOVERY_LATENCY}] {data/export/vanilla_auth.csv};
        \addplot+ [boxplotcolor=CoreBlue, boxplot, draw=black] table [col sep={comma},y={DISCOVERY_LATENCY}] {data/export/dns_discovery.csv};
        \addplot+ [boxplotcolor=CoreGreen, boxplot, draw=black] table [col sep={comma},y={DISCOVERY_LATENCY}] {data/export/dns_and_dane.csv};
        \end{axis}
    \end{tikzpicture}
    \vspace{-15pt}
    \caption{
        Whisker plot on the latency of service discovery.
    }
    \label{fig:sdlatency}
% \end{figure}
    \end{minipage}%
    \hspace{12pt}%
    \begin{minipage}[t]{.485\textwidth}
        %!TEX root = ../../main.tex

% \begin{figure}
    % \centering
    \begin{tikzpicture}
        \begin{axis}[
            width=\linewidth,height=.8\linewidth,
            boxplot/draw direction=y,
            xtick = {1,2,3,4},
            xticklabels = {SOME/IP SD, SOME/IP SD\\ w/ AUTH, DNSSEC, DNSSEC \\w/ DANE},
            xticklabel style = {anchor=north east,align=center, font=\small, rotate=35},
            ylabel = {Subscription latency},
            y unit = {\milli\second},
            ymax = 10,
            enlarge y limits,
            ymajorgrids,
            % xtick style = {draw=none}, % Hide tick line
            boxplotcolor/.style={mark=square*, color=#1,fill=#1,mark options={color=#1,fill=#1,draw=black}},
            ]
        \addplot+ [boxplotcolor=CoreRed, boxplot, draw=black] table [col sep={comma},y={SUBSCRIPTION_LATENCY}] {data/export/vanilla_no_auth.csv};
        \addplot+ [boxplotcolor=CoreYellow, boxplot, draw=black] table [col sep={comma},y={SUBSCRIPTION_LATENCY}] {data/export/vanilla_auth.csv};
        \addplot+ [boxplotcolor=CoreBlue, boxplot, mark=none, draw=black] table [col sep={comma},y={SUBSCRIPTION_LATENCY}] {data/export/dns_discovery.csv};
        \addplot+ [boxplotcolor=CoreGreen, boxplot, draw=black] table [col sep={comma},y={SUBSCRIPTION_LATENCY}] {data/export/dns_and_dane.csv};
        \end{axis}
    \end{tikzpicture}
    \vspace{-15pt}
    \caption{
        Whisker plot on the latency of service subscriptions. The small squares are outliers.
        }
    \label{fig:sublatency}
% \end{figure}
    \end{minipage}%
\end{figure*}

\subsection{Evaluation Setup}
We measure the service discovery and subscription latency and the cost of the cryptographic operations.
We do not directly compare our solutions to~\cite{irrsvssvJR20,zlkkassJR21,myzzcascJR22} since we do not have access to their implementations, but we use similar cryptographic operations for authentication.
Compared to pre-deployed certificates~(cf.~\cite{irrsvssvJR20,zlkkassJR21}), the \ac{DNS} resolver is introduced as an additional instance~(cf.~\cite{zlkkassJR21,myzzcascJR22}) with \ac{DNSSEC} and the data authenticity is outsourced to the~\ac{DNSSEC}.
In doing so, we compare four different solutions, all of which are implemented based on the \textit{vsomeip}\cite{vsomeip-git} stack:
\begin{enumerate}
    \item \textbf{SOME/IP SD:} An unaltered \textit{vsomeip} implementation that we use as a baseline.
    \item \textbf{DNSSEC}: A \ac{DNSSEC} augmented \textit{vsomeip} that replaces the original \codeword{offer}/\codeword{find} procedure with our \ac{DNSSEC}-based consumer-triggered discovery.
    \item \textbf{DNSSEC w/ DANE} An authentication approach based on the \ac{DNSSEC} discovery implementation that uses \ac{DANE} records to retrieve the publisher certificate, and our challenge-response mechanism during the subscription phase to authenticate the publisher.
    \item \textbf{SOME/IP SD w/ AUTH}: An authenticated approach returning to the original \ac{SOME/IP} \ac{SD} without any \ac{DNS} operations that uses pre-deployed certificates and our challenge-response mechanism to authenticate the publisher.
\end{enumerate}

Our evaluation setup consists of three nodes for a client, a server and a DNSSEC resolver, which are arranged as shown in Figure~\ref{fig:someipsddnsmod}.
All nodes run on the same host system (CPU: AMD FX-8350 with 8 cores at 4Ghz, RAM: 16GB) in separate containers (\textit{Docker}: 20.10.22) connected via the Docker virtual bridge network.

The server and client containers run on a Linux OS with the \ac{SOME/IP} stack, and libraries for \ac{DNS} lookups and cryptography 
(\textit{Ubuntu}: 18.04.6 LTS, 
\textit{vsomeip}~\cite{vsomeip-git}: 3.1.20.3, 
\textit{Crypto++}~\cite{cryptopp-website}: 8.7.0, 
\textit{Crypto++ PEM Pack}: 8.2).
The \ac{DNSSEC} resolver runs on a Linux OS (\textit{Ubuntu}: 22.04.1 LTS, \textit{Unbound}: 1.17.0).

The DNS entries for the SVCB and TLSA records of the publisher service are already in the cache of the DNSSEC resolver and validated along the trust chain, as would be the case in an automotive deployment. 

As \ac{SOME/IP} uses group communication for service discovery, it applies common practices for scattering multicast communication to reduce the load on the network and hosts.
For example, responses can be delayed collecting multiple requests and answer them in a single response, and a random initial delay prevents all \acp{ECU} from flooding the network by sending discovery messages at the same time.
Since we compare it to standard \ac{DNS} discovery via unicast queries, which does not include any of such delays, we turn off the request-response delay in \textit{vsomeip} to get comparable results.
Moreover, we only look at the connection of one server and client, for which these mechanisms are not needed. 
However, the startup phases of \ac{SOME/IP} \ac{SD} remain unchanged, and thus a random initial delay between \SI{10}{\milli\second} and \SI{100}{\milli\second} delays the startup of the discovery phase.

\subsection{Discovery and Subscription Latency Benchmark}
Our benchmark evaluates the latencies of the discovery, subscription, and cryptographic operations. 
For each of the four compared solutions, we collect fifty samples with timestamps indicating the beginning and end of different phases to calculate the latency based on the difference between these timestamps.
Figure~\ref{fig:someipsd} and Figure~\ref{fig:someipsdaugdns} show the sequences of the consumer-triggered discovery and publish-subscribe phases for the \ac{SOME/IP} \ac{SD} and the \ac{DNS} discovery, respectively.

Figure~\ref{fig:sdlatency} shows the consumer-triggered discovery latency of all four different solutions.
Here, we measure the time that elapses from the completion of the initialization of the client until the result of the service discovery is available.
Since the measured interval for the discovery does not include authentication operations, the latency of the solutions with publisher authentication are expected to be the same as without publisher authentication.
The \ac{DNS} discovery latencies are between \SI{4}{\milli\second} and \SI{6}{\milli\second}.
Both \ac{SOME/IP} \ac{SD} variants have a latency between \SI{13}{\milli\second} and \SI{103}{\milli\second} due to the random initial delay between \SI{10}{\milli\second} and \SI{100}{\milli\second}.
Without an initial delay, the latency of the \ac{SOME/IP} \ac{SD} would be similar to that of the \ac{DNS} discovery.

Figure~\ref{fig:sublatency} shows the subscription latency of the four candidates.
We measure the time that elapses between the sending of the first subscription message and the completion of the connection setup, including the verification of the publisher signature in the authenticated approaches.
The solutions without publisher authentication have a latency under \SI{1}{\milli\second}.
With publisher authentication the latency is between \SI{4}{\milli\second} and \SI{9}{\milli\second}.
In detail, signing the nonce at the publisher takes between \SI{3}{\milli\second} and \SI{7}{\milli\second}, verifying the signature at the client side is below \SI{2}{\milli\second}.
The trade-off in using our challenge-response scheme results in a maximum delay of \SI{8}{\milli\second}.

Considering the overall discovery and subscription latency the publisher authentication does not have a significant impact on the latency, for which the multicast scattering is the most notable delay.
\ac{DNSSEC} and \ac{DANE} enable publisher authenticity without a large performance penalty even compared to authentication with pre-deployed certificates.
We achieve this by querying the TLSA record at the same time as we initiate the subscription.
However, the latency of the TLSA \codeword{response} containing the certificate depends on the link to the \ac{DNS} server.
This could impact the results when the DNS \codeword{query} takes longer than the subscription handshake.
Here, an evaluation with an Ethernet-connected \ac{DNS} server would be interesting to see the impact of the latency.
In addition, the performance for a larger number of services should be analyzed to determine scalability with \ac{DNS} discovery compared to \ac{SOME/IP} \ac{SD} in a realistic automotive network.

%!TEX root = ../main.tex

\section{Conclusion and Outlook}%
\label{sec:conclusion_and_outlook}
In this paper, we designed and analyzed basic security elements for the rapidly evolving service-oriented software architecture in future cars. In provisioning service authentication and managing attestation credentials, we  addressed the urgent demand for securing a heterogeneous, distributed, and dynamically updatable software ecosystem that will drive the connected cars of the near future.

Our work was intentionally built on well-established standards. \ac{DNSSEC} and \ac{DANE}  enable certificate management and service authenticity while being a thoroughly validated, operationally stable Internet standard. \ac{SOME/IP} is a widely accepted service-oriented middleware standardized by AUTOSAR. We demonstrated how to combine  the \ac{SOME/IP} \ac{SD} with the Internet name system in design, implementation, and evaluation. Our findings indicated that \ac{SOME/IP} \ac{SD} can interact with the DNS without operational overhead, while  \ac{DNSSEC} with \ac{DANE} contribute not only a robust, reliable security solution but also a stable infrastructure for replication, (off-line) caching, and key management.

This basic solution to automotive service security opens three future research directions. First, the remaining \ac{SOME/IP} service primitives for onboard session establishment and migration need a detailed  security design and assessment. Second, operational guidelines for namespace management and service updates in the automotive ecosystem shall be developed. Third, we aim at configuring a full-featured production-grade vehicle with our security solution and evaluate its properties in macroscopic benchmarks.

%\newpage

%%%% 	BibTeX		%%%%
%\bibliographystyle{plain}
\bibliographystyle{IEEEtran}
\bibliography{bibtex/HTML-Export/all_generated,bib/own,bib/security,bib/rfcs,bib/literature,bib/bibliography}

\end{document}